\documentclass[final]{svjour3}
\usepackage{graphicx}
\usepackage{rotating}
\usepackage{amssymb}
\usepackage{mathptmx}
\usepackage[square,sort,comma,numbers]{natbib}
\usepackage[colorlinks=true, allcolors=blue]{hyperref}
\makeatletter
\journalname{Journal of Low Temperature Physics}

\begin{document}

\newcommand{\umux}{$\mu$MUX }
\newcommand{\sups}[1]{\textsuperscript{#1}}
\newcommand{\hdblarrow}{H\makebox[0.9ex][l]{$\downdownarrows$}-}
\title{The Design of The CCAT-prime Epoch of Reionization Spectrometer Instrument}

\author{N.F. Cothard\sups{1} \and
S.K. Choi\sups{2} \and
C.J. Duell\sups{3} \and
T. Herter\sups{2} \and
J. Hubmayr\sups{4} \and
J. McMahon\sups{5} \and
M.D. Niemack\sups{3} \and
T. Nikola\sups{6} \and
C. Sierra\sups{5} \and
G.J. Stacey\sups{2} \and
E.M. Vavagiakis\sups{3} \and
E.J. Wollack\sups{7} \and
B. Zou\sups{1}
}

\institute{
\sups{1}Department of Applied Physics, Cornell University, Ithaca, NY 14853, USA\\
\sups{2}Department of Astronomy, Cornell University, Ithaca, NY 14853, USA\\
\sups{3}Department of Physics, Cornell University, Ithaca, NY 14853, USA\\
\sups{4}NIST Quantum Devices Group, Boulder, CO 80305, USA\\
\sups{5}Department of Physics, University of Michigan, Ann Arbor, MI 48103, USA\\
\sups{6}Cornell Center for Astrophysics and Planetary Science, Cornell University , Ithaca, NY 14853, USA\\
\sups{7}NASA Goddard Space Flight Center, Greenbelt, MD 20771, USA\\
\email{nc467@cornell.edu}}

\maketitle

\begin{abstract}

The Epoch of Reionization Spectrometer (EoR-Spec) is an instrument module for the Prime-Cam receiver of the 6 m aperture CCAT-prime Telescope at 5600 m in Chile. EoR-Spec will perform 158 $\mu$m [CII] line intensity mapping of star-forming regions at redshifts between 3.5 and 8 (420 - 210 GHz), tracing the evolution of structure during early galaxy formation. At lower redshifts, EoR-Spec will observe galaxies near the period of peak star formation - when most stars in today's universe were formed. At higher redshifts, EoR-Spec will trace the late stages of reionization, the early stages of galaxy assembly, and the formation of large-scale, three-dimensional clustering of star-forming galaxies. To achieve its science goals, EoR-Spec will utilize CCAT-prime's exceptionally low water vapor site, large field of view ($\sim 5$ degrees at 210 GHz), and narrow beam widths ($\sim 1$ arcminute at 210 GHz). EoR-Spec will be outfitted with a cryogenic, metamaterial, silicon substrate-based Fabry-Perot Interferometer operating at a resolving power ($\lambda/\Delta\lambda$) of 100. Monolithic dichroic arrays of cryogenic, feedhorn-coupled transition edge sensor bolometers provide approximately 6000 detectors, which are read out using a frequency division multiplexing system based on microwave SQUIDs. The novel design allows the measurement of the [CII] line at two redshifts simultaneously using dichroic pixels and two orders of the Fabry-Perot. Here we present the design and science goals of EoR-Spec, with emphasis on the spectrometer, detector array, and readout designs.

\end{abstract}

\section{Introduction}

The epoch of reionization (EoR) is a largely unexplored cosmic epoch ($\sim6 < \textrm{z} < 11$) during which the neutral hydrogen that formed during the epoch of recombination began to be re-ionized by the ultraviolet light of the first stars and/or black hole accretion activity. 
Studies of individual, bright, high-redshift galaxies with HST and ALMA hint that cosmic reionization is driven by the starlight from stars within the first galaxies that form in the universe \cite{riechers_alma_2014}.
However, the process of reionization remains poorly understood since it is thought to be driven by a multitude of intrinsically faint sources.
Individually these faint sources are very difficult to detect but in aggregate their emissions are detectable by instruments having high surface brightness sensitivity.
Spectral line imaging of the aggregate signal, often called line intensity mapping (LIM), overcomes this challenge by measuring the spatial fluctuations of large-scale structure spectroscopically at low spatial resolution \cite{kovetz_line-intensity_2017}.
By mapping sources of emission or absorption in three dimensions (as a function of z), LIM can be used to understand the processes of structure formation during the EoR.
The 158 $\mu$m [CII] cooling emission of star forming regions is an exceptionally bright and efficient tracer of early structure formation and reionization \cite{stacey_158_1991,stacey_158_2010}.
Our EoR Spectrometer (EoR-Spec) will perform such intensity mapping measurements of the [CII] line with the combination of a high-throughput sub-millimeter telescope, cryogenic interferometer, and superconducting, photon-noise limited detector arrays.

In Section 2 we present the science objectives of EoR-Spec and in Section 3 we discuss the instrument design to achieve these objectives. We conclude in Section 4 with an update on the status and plans for the experiment.

\section{Science Objectives and Motivation}

EoR-Spec will explore the Epoch of Reionization via the cooling radiation in the 158 $\mu$ m [CII] line from star forming galaxies.
The first galaxies form at dark matter overdensities such that their ionizing starlight illuminates the underlying dark matter structure of the early universe \cite{kovetz_line-intensity_2017}.
Through LIM, EoR-Spec will investigate the process of structure formation, probing the collective star formation from galaxies and the underlying large-scale matter density fluctuations.

\begin{figure}[htbp]
\begin{center}
\includegraphics[width=\linewidth, keepaspectratio]{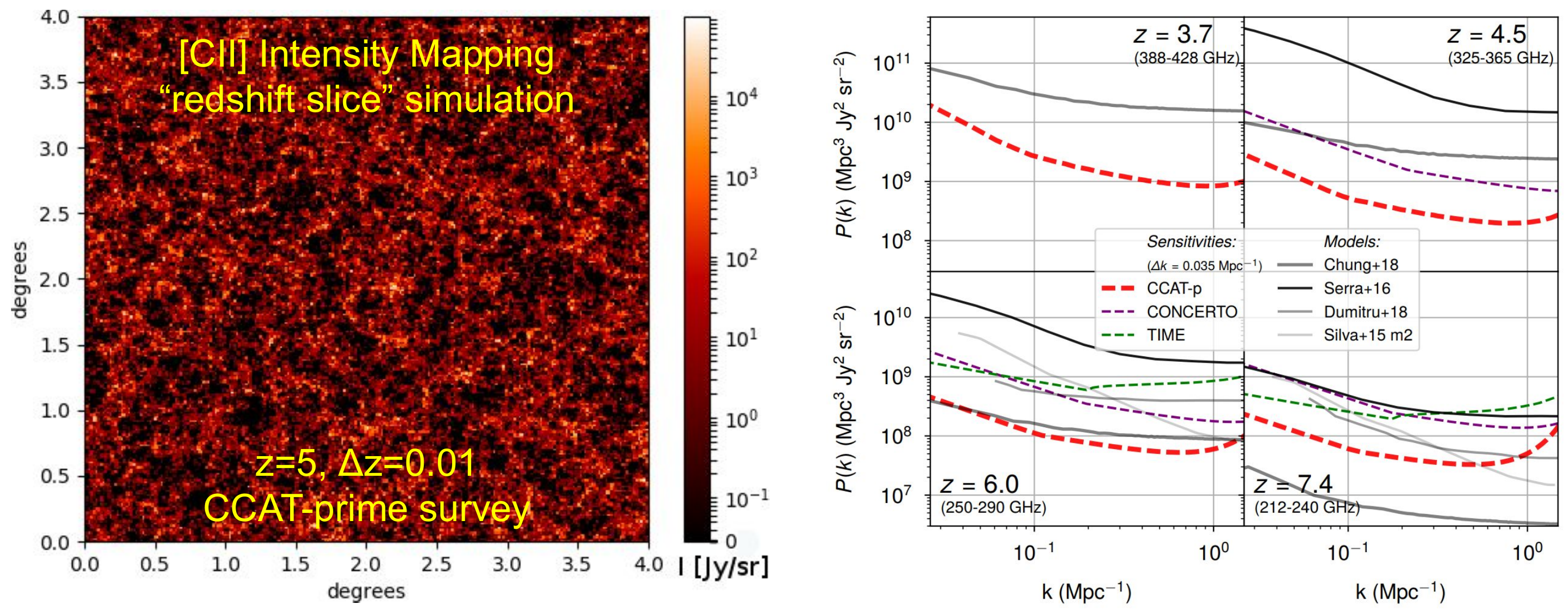}
\caption{
{\it Left}: simulated redshift slice of [CII] emission at $\textrm{z} \sim 5$ from a 16 deg\sups{2} survey showing the clustering signal \cite{slice}.
Represented is a single spectral bin of the spatial-spectral intensity mapping data cube.
{\it Right}: Sensitivity curves for the power spectrum of the [CII] emission at multiple redshifts \cite{sensitivity}.
Plotted are multiple power spectra predictions which differ by factors of 10 to 50. Over plotted are predicted EoR-Spec sensitivity curves for the first generation 16 deg\sups{2} survey as well as sensitivity curves for other upcoming experiments.
}
\label{fig1}
\end{center}

\end{figure}

This survey of [CII] emission from the EoR will be highly complementary with other intensity mapping probes such as HI 21 cm, H$\alpha$, H$\beta$, OIII, Ly$\alpha$, and CO \cite{kovetz_line-intensity_2017,dore_cosmology_2014}.  
For example, the HI 21 cm line emission from neutral hydrogen is expected to anti-correlate with [CII], since the [CII] line will trace gas heated, or ionized by starlight.
Correlating the brightness and three-dimensional spatial distributions of different probes will reveal properties of the radiation fields, such as the numbers, types, and spatial distribution of young massive stars within ionizing galaxies. 
Furthermore, the combined probes will explore the growth of ionization, the production of metals, and the evolution of reionization bubbles \cite{kovetz_line-intensity_2017}.
 
By intensity mapping the aggregate [CII] emission over large co-moving spatial scales, EoR-Spec will be a cosmological probe, revealing the growth of density fluctuations and the primordial power spectrum (Figure \ref{fig1}, {\it Right}). 
With sufficiently precise measurements, constraints could be placed on fundamental physics including models of inflation, dark energy, and the sum of the neutrino masses \cite{kovetz_line-intensity_2017}.
While some of these results could be obtained through pointed measurements of individual galaxies, intensity mapping provides a much more efficient measurement.
Rather than using the largest aperture telescopes and very long integration times for individual sources, EoR-Spec will be sensitive to the aggregate signal of all bright and faint sources.
This large-scale mapping effort will be enabled by CCAT-prime's modest 6 m aperture and very large field of view.
EoR-Spec's intensity maps will therefore track the evolution of star and galaxy formation, including diffuse emission from dim dwarf galaxies and the brightest galaxies. 
Preliminary plans for survey areas and sensitivity estimates can be found in \cite{stacey_ccat-prime:_2018,steve_ltd2019}.

\section{Instrument Overview}
EoR-Spec will be a spectrometer instrument module in CCAT-prime's Prime-Cam receiver, optimized for intensity mapping the redshifted [CII] line between 210 and 420 GHz.
CCAT-prime is a 6 m aperture telescope being constructed at 5600 m on Cerro Chajnantor in northern Chile \cite{parshley_ccat-prime:_2018}.
With its exceptionally dry site, CCAT-prime is designed to take advantage of the extremely low water vapor in the millimeter and sub-millimeter bands \cite{stacey_ccat-prime:_2018}.
EoR-Spec will occupy one instrument module in the Prime-Cam receiver, which will provide continuous cooling power with a dilution refrigerator, keeping the detector arrays at 100 mK \cite{vavagiakis_prime-cam:_2018}.
For the EoR-Spec module, CCAT-prime will provide a wide 1.3 degree field of view and diffraction limited beam of 30-60 arcseconds (420-210 GHz, respectively), which is well matched to the Mpc clustering scale of the EoR signal.

EoR-Spec will use a silicon substrate-based, scanning Fabry-Perot interferometer (FPI) to illuminate the broadband detectors with narrow bandpasses. 
The interferometer will be installed at the Lyot stop of the optics tube, which will be cooled to 4K with a pulsetube cryo-cooler.
The optics of the instrument module are optimized to provide well-collimated beams at the Lyot stop.
The scanning FPI will be used to shift the resonant FPI bandpasses across the the broadband detectors, enabling spectroscopic measurements.
The silicon substrate FPI will be comprised of two silicon mirrors which will be patterned with highly reflective metal meshes on one side, and broadband metamaterial anti-reflection coatings (ARCs) on the other.
EoR-Spec is baselining a detector array of feedhorn-coupled, dichroic, transition edge sensor (TES) pixels.
With a dichroic detector array, EoR-Spec will observe two orders of the FPI simultaneously, increasing mapping speed by up to a factor of two.

\subsection{\it Optics}

\begin{figure}[htbp]
\begin{center}
\includegraphics[width=\linewidth, keepaspectratio]{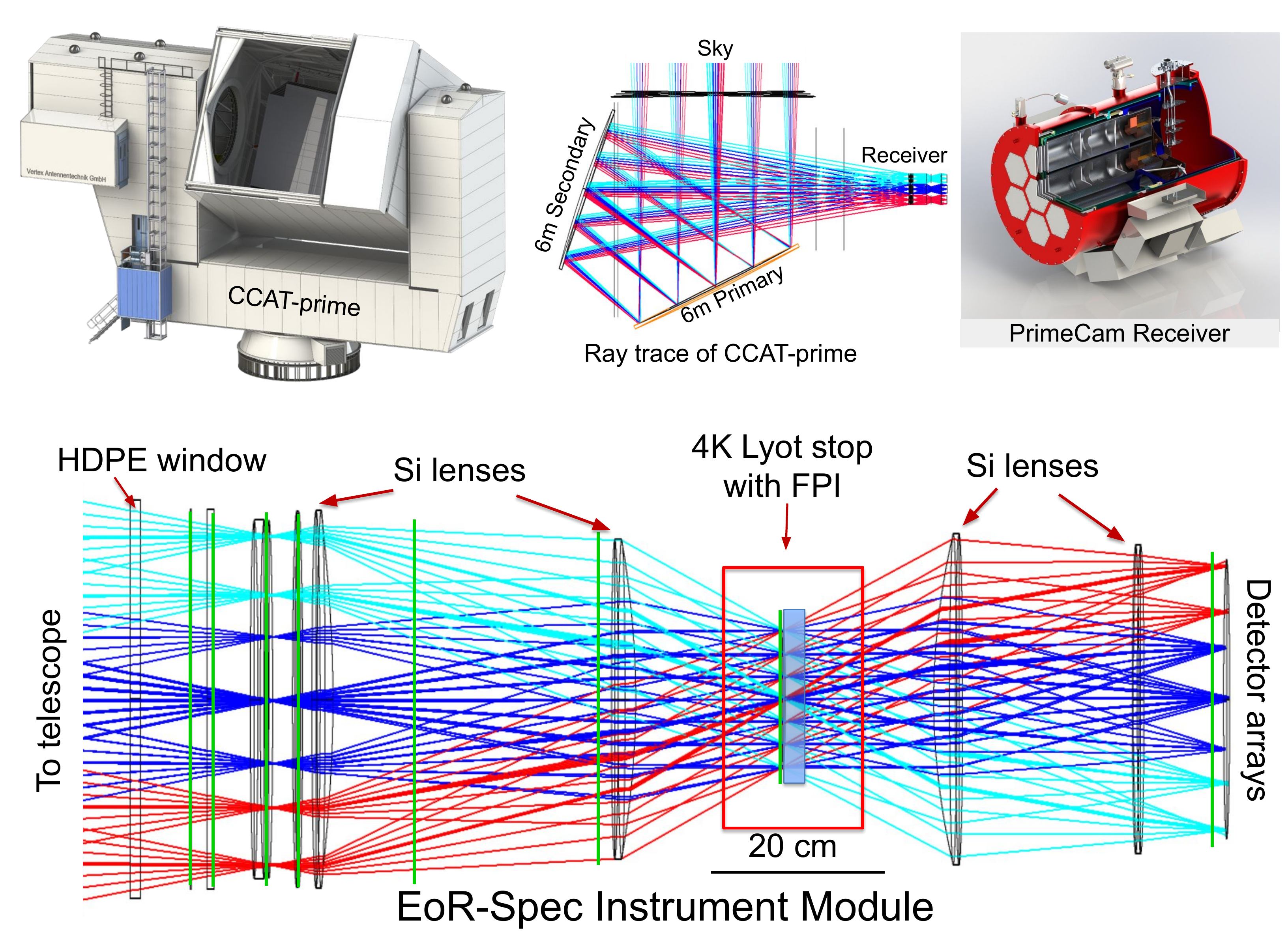}
\caption{
Renderings of the CCAT-prime telescope ({\it top left})and Prime-Cam receiver ({\it top right}), and ray traces of the telescope ({\it top center}) and EoR-Spec instrument module ({\it bottom}).
The instrument module uses anti-reflection coated silicon lenses to re-image the sky onto the detector arrays.
The optics of the module have been optimized to maximize beam collimation through the Fabry-Perot interferometer.
}
\label{fig2}
\end{center}

\end{figure}

The Prime-Cam receiver will sit at the f/2.7 focus of CCAT-prime, with a 7.8 degree field of view, extending over 2 m in diameter in the image plane. 
The EoR-Spec instrument module will occupy 1.3 degrees of the image plane. 
Figure \ref{fig2} shows renderings of the CCAT-prime telescope and Prime-Cam receiver, and ray traces of the telescope and EoR-Spec instrument module.
The entrance window will be high-density polyethylene with a laminated, expanded, teflon, anti-reflection coating.
Silicon lenses with metamaterial ARCs create collimated beams passing through the 4 K FPI at the Lyot stop and subsequently focus the beams onto the detector arrays.
Metal mesh IR-blocking and low-pass filters will be used throughout to minimize out of band thermal and detector loading.

\subsection{\it Fabry-Perot Interferometer}

\begin{figure}[htbp]
\begin{center}
\includegraphics[width=1\linewidth, keepaspectratio]{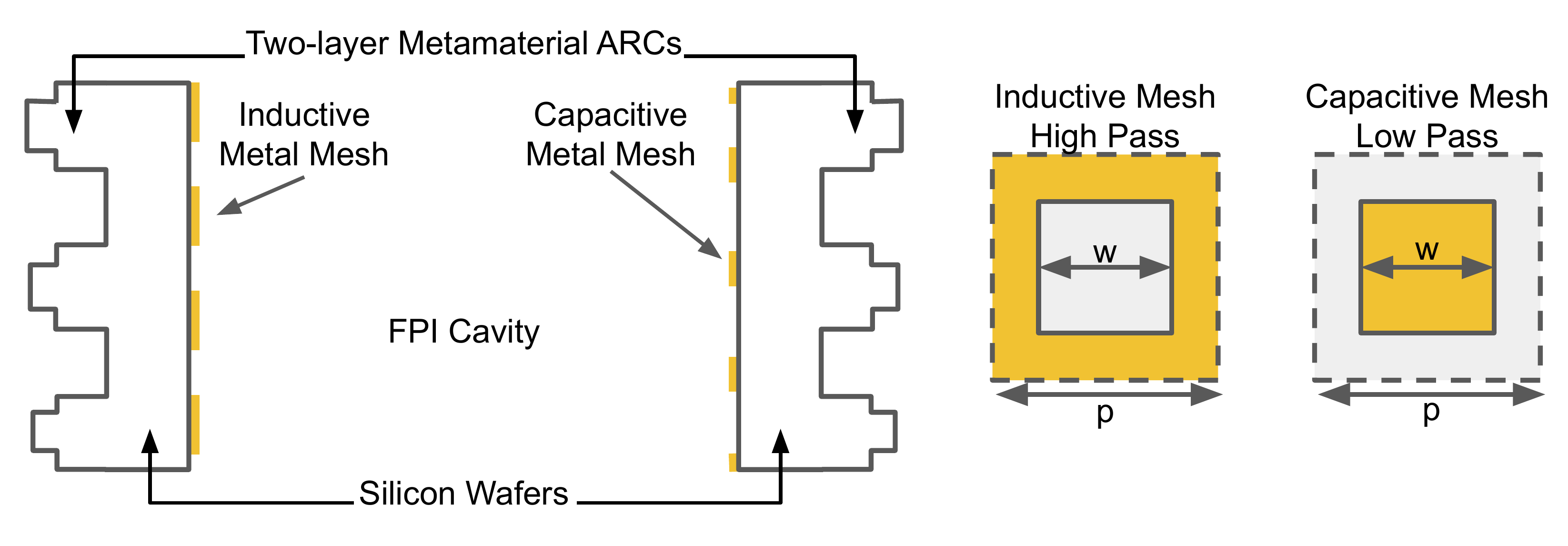}
\caption{
An illustration of the SSB FPI consisting of two silicon wafers with two-layer metamaterial ARCs and inductive and capacitive metal mesh reflectors. The diagram is not to scale. To place the second order resonance at $\sim300$ GHz, the mirror separation will be $\sim 1$ mm. The combination of inductive and capacitive meshes flattens the finesse of the FPI across a wider bandwidth \cite{cothard_optimizing_2018}.
}
\label{fig3}
\end{center}
\end{figure}

\begin{figure}[htbp]
\begin{center}
\includegraphics[width=\linewidth, keepaspectratio]{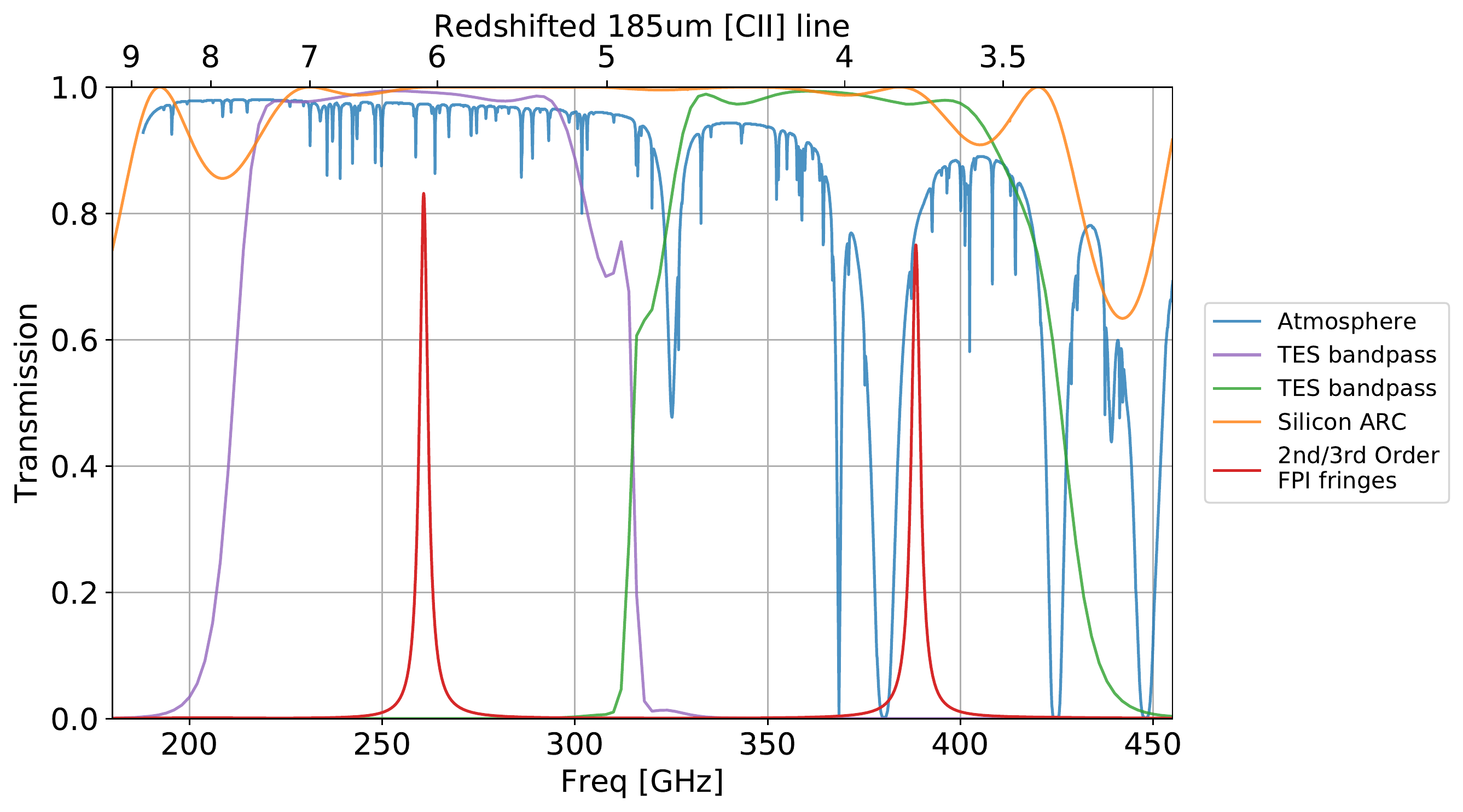}
\caption{
Transmission profile of the interferometer. 
Over plotted is the zenith transmission in the first quartile of the atmosphere at the CCAT-prime site (blue), modeled two layer metamaterial antireflection coating (orange), modeled dichroic TES pixel bandpasses (purple and green), and a model of the Fabry-Perot Interferometer (red).
The second and third order FPI resonances are loading different TESes simultaneously, providing spectral multiplexing.
}
\label{fig4}
\end{center}
\end{figure}

A scanning, cryogenic Fabry-Perot interferometer (FPI) will enable our intensity mapping observations by providing high-sensitivity, wide-field, broad-band spectroscopy.
During observations, the telescope will be scanned across the sky with the FPI in a fixed position. 
After several sky scans, the FPI will be stepped, shifting the resonant frequencies of the interferometer, and then scanning will be repeated.
The spatial-spectral data cube is constructed by spatially mapping the sky at each step of the FPI's frequency range until the entire spectrum is obtained to the required depth.

The optical design of EoR-Spec is optimized for the FPI by providing well collimated beams at the Lyot stop.
The FPI consists of two 200 mm diameter silicon substrate based (SSB) mirrors that form a resonating cavity.
SSB mirrors offer many advantages over traditional freestanding metal mesh mirrors such as greater flexibility over the mirror reflectivity, control over ohmic losses, control over surface roughness, mechanical stability, and thermal performance \cite{cothard_optimizing_2018}. 

SSB mirror fabrication is enabled by microfabrication techniques adapted from the silicon industry.
The mirrors are comprised of high-purity, float zone, 500 $\mu$m thick silicon wafers which are lithographically patterned with a frequency-selective, reflective gold mesh on one side \cite{ade_review_2006}.
Due to the high index of refraction of silicon (n $\sim 3.4$), the other side of the mirror must be patterned with an ARC.
Deep reactive ion etching (DRIE), a plasma etching process, is used to create a two-layer metamaterial ARC \cite{gallardo_deep_2017,cothard_optimizing_2018}.
A two-layer ARC is necessary in order to provide adequate suppression of parasitic reflections across the wide bandwidth of the instrument.
The metamaterial coating enables precise ARC optimization while also keeping the ARC thermally matched to the FPI substrate.
Figure \ref{fig3} is an illustration of the SSB FPI, consisting of a pair of silicon wafers with double layer ARCs and inductive and capactive metal mesh reflectors.
Figure \ref{fig4} shows the modeled transmission of the SSB FPI and metamaterial ARC, as well as the atmospheric transmission (zenith, first quartile) and the TES detector broadband bandpasses.

The SSB FPI is a novel application of previously fielded technologies, with strong heritages. 
Silicon-based metamaterial ARCs in the millimeter and sub-millimeter are being commonly used in cosmic microwave background experiments such as AdvACT and Simons Observatory, where multi-layer ARCs have been cut into the surface of silicon lenses using dicing saws \cite{datta_large-aperture_2013}.
Lithographically patterned metal meshes are used widely in the far-IR to millimeter as IR-rejection and band-defining filters \cite{ade_review_2006} using.
These technologies have been combined in the past to produce silicon substrate-based IR blocking filters by sandwiching low-pass metal mesh filters with IR absorbing material in between to AR coated silicon wafers \cite{munson_composite_2017}.
Here, we have discussed the novel application of these technologies for use in high performance, wide bandwidth Fabry-Perot interferometer cavities.
\subsection{\it Detectors}

\begin{figure}[htbp]
\begin{center}
\includegraphics[width=\linewidth, keepaspectratio]{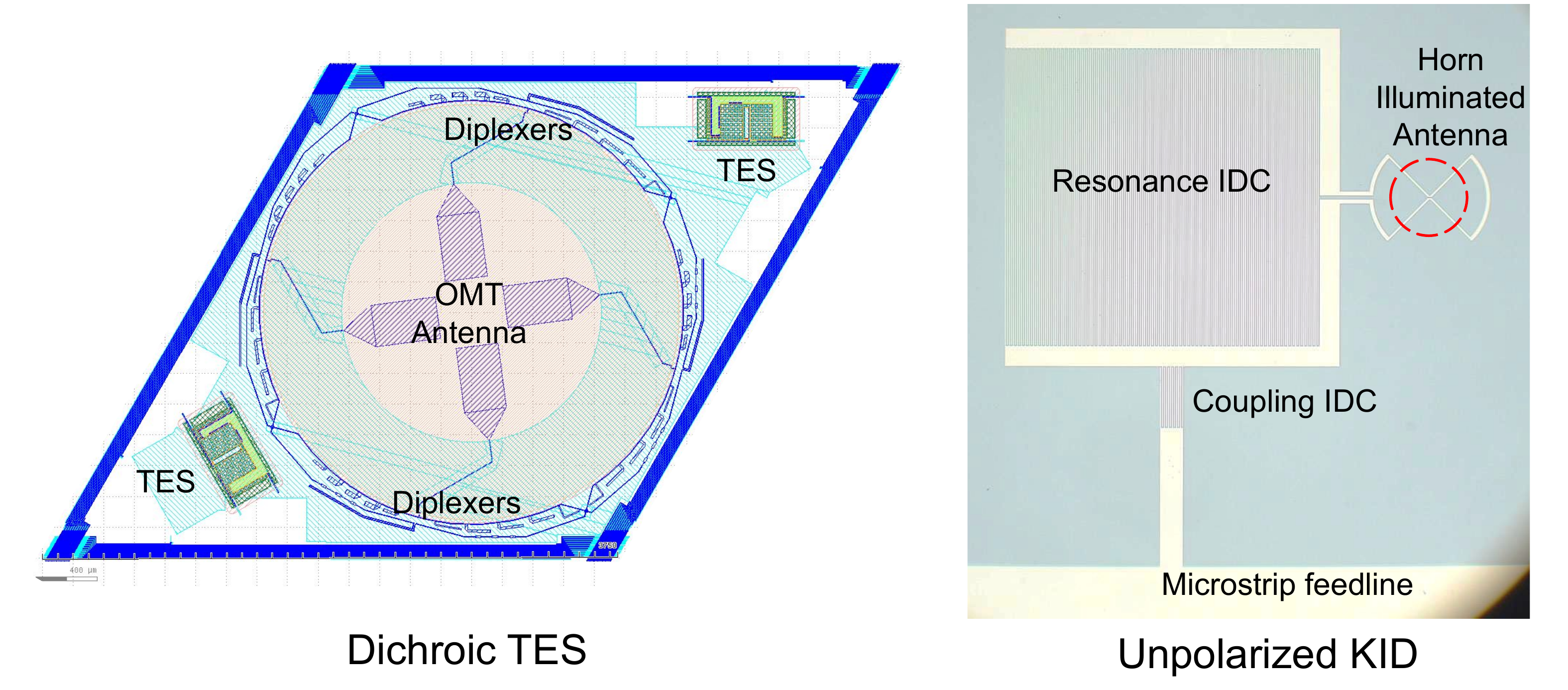}
\caption{
{\it Left}: Layout of a dichroic TES pixel concept showing the OMT antenna at the center of the pixel, frequency diplexing microwave stub filters, and TES bolometer islands for two frequency bands. The parallelogram outline of the pixel is comprised of superconducting wire busses that connect the TES to a readout system. For each color, both polarizations are terminated onto a single TES bolometer.
{\it Right}: Micrograph of an unpolarized KID with horn coupled antenna. Many KIDs are coupled to a common microstrip feedline for readout.
}
\label{fig5}
\end{center}
\end{figure}

Beams through the FPI are focused onto three feedhorn-coupled detector arrays, fabricated on 150 mm diameter silicon wafers.
EoR-Spec is baselining transition edge sensor (TES) bolometer arrays of $\sim2000$ detectors per wafer, fabricated at NIST in Boulder, Colorado.
The TES pixel design is adapted from previously fielded and successful instruments, such as Advanced ACTPol \cite{henderson_advanced_2016,choi_characterization_2018,koopman_advanced_2018}.
For each feedhorn-coupled pixel in the array, there are two TES detectors, one for each frequency band (see Figure \ref{fig5}, {\it Left}).
Each pixel uses an orthomode transducer antenna to couple each polarization onto planar, superconducting, on-chip microwave circuitry.
Each polarization is fed to a diplexer which defines two frequency bands, approximately 210-315 GHz and 315-420 GHz (See Figure \ref{fig4}). 
For each frequency band, there is a single TES bolometer where power from both polarizations are terminated.
The geometry of the bolometer islands and absorbers will be tuned for the low loading conditions of the spectrometer.
Expected optical loadings are in the $\sim 1 ~$pW range.

The combination of the imaging FPI and a dichroic detector array provides a well spectrally and spatially multiplexed system to recover the faint large-scale signals inherent in intensity mapping.
In particular, dichroic pixels allow the FPI to be operated with two transmitting modes. 
The second and third order modes will be simultaneously observed with the low and high frequency detectors respectively (see Figure \ref{fig4}), increasing mapping speeds by up to a factor of two.
Additionally, off-axis detectors in the focal plane go through the FPI at different angles, resulting in bluer resonant orders than on-axis pixels.
As a result, scanning across the sky naturally provides simultaneous spectral and spatial multiplexing, even at a fixed FPI setting.

While TES arrays are the baseline technology, the EoR-Spec team is also investigating the use of kinetic inductance detector (KID) arrays.
With background-limited detectors, the mapping speed is limited by the number of detectors that can be readout in each array.
A KID array, which couples large numbers of KIDs to a common feedline (see Figure \ref{fig5}, {\it Right}) in EoR-Spec could contain up to twice the number of detectors as the baseline TES array.
If a KID array can be optimized for the low loading conditions of the spectrometer, then the mapping speed could be doubled.

\subsection{\it Readout}

Each TES in the baseline EoR-Spec arrays will be DC voltage biased and read out using microwave SQUID multiplexing (\umux), a frequency division multiplexing (FDM) system that inductively couples each TES to a unique microresonator through an rf-SQUID \cite{dober_microwave_2017}.
Each unique microresonator is then weakly coupled to a common microwave feedline.
This readout system is based on \umux  systems being developed and deployed for the BICEP-Keck array and Simons Observatory, in which thousands of TES-coupled resonators are being readout simultaneously.
Prime-Cam will read out many \umux  arrays using the SLAC Microresonator Radio Frequency (SMuRF) electronics \cite{kernasovskiy_slac_2018,max_ltd2019}. The SMuRF system enables readout of microresonator-coupled bolometers by implementing TES bias lines, rf-SQUID flux ramp modulation, and closed loop resonator tone tracking.

\section{Status}

EoR-Spec will be one of the first instruments in Prime-Cam on CCAT-prime, which will finish construction in 2021. 
Development and fabrication of the FPI and detector arrays is underway, with testing planned for 2020 and 2021.

\begin{acknowledgements}
Work by NFC was supported by a NASA Space Technology Research Fellowship.
Fabrication work at the Cornell NanoScale Science and Technology Facility supported by NASA Grant NNX16AC72G.
MDN acknowledges support from NSF award AST-1454881. 
This work was performed in part at Cornell NanoScale Facility, an NNCI member supported by NSF Grant ECCS-1542081.
\end{acknowledgements}


\bibliographystyle{ltd_style}
\bibliography{references,extra}



\end{document}